\documentclass[runningheads]{llncs}
\usepackage{graphicx}
\usepackage{hyperref}
\usepackage{amsmath, amssymb, mathtools} 

\def\*#1{\mathbf{#1}}
\def\^#1{\boldsymbol{#1}}

\begin{document}
\title{Inference in neural networks using conditional mean-field methods}
%
%
\author{Ángel Poc-López\inst{1,\ddag,*}\orcidID{0000-0001-5104-6963} \and
\\  Miguel Aguilera\inst{2,1,\ddag,*}\orcidID{0000-0002-3366-4706} }
\authorrunning{A. Poc-López and M. Aguilera}
%
\institute{ISAAC Lab, I3A Arag\'on Institute of Engineering Research, University of Zaragoza, Zaragoza, Spain \and
Department of Informatics  Sussex Neuroscience, University of Sussex, Brighton, UK 
\\ $^\dag$ \email{angel.poc.lopez@gmail.com} 
\\ $^\ddag$ \email{sci@maguilera.net}
\\$^*$  Both authors contributed equally to this work}
%
\maketitle              
\begin{abstract}
We extend previous mean-field approaches for non-equilibrium neural network models to estimate correlations in the system. This offers a powerful tool for approximating the system dynamics as well as a fast method to infer network parameters from observations. We develop our method in an asymmetric kinetic Ising model and test its performance on 1) synthetic data generated by an asymmetric version of the Sherrington Kirkpatric model and 2) recordings of  in vitro neuron spiking activity from the mouse somatosensory cortex. We find that our mean-field method outperforms previous ones in estimating networks correlations and successfully reconstructs network dynamics from data near a phase transition showing large fluctuations.

\end{abstract}

\section{Introduction}

Biological and neural networks generally exhibit out-of-equilibrium dynamics \cite{nicolis_self-organization_1977}. Resulting physiological rhythms and emerging patterns are in continuous and asymmetrical interactions within and between networks. Furthermore, such networks are often found to self-organize near critical regimes at which their fluctuations are maximized \cite{tkacik_thermodynamics_2015}.
Although new data acquisition technologies are providing detailed descriptions of the dynamics of hundreds or thousands of neurons in different animals \cite{ahrens_whole-brain_2013,stringer_high-dimensional_2019}, these properties make it challenging to analyze the evolution of such systems assuming an asymptotic equilibrium state or standard approximation methods. This problem demands mathematical tools for capturing and reproducing the types of non-equilibrum fluctuations found in large biological systems.

The kinetic Ising model with asymmetric couplings is a prototypical model for studying such non-equilibrium dynamics in biological systems \cite{roudi_multi-neuronal_2015}.
The model is described as a discrete-time Markov chain of interacting binary units, resembling the nonlinear dynamics of recurrently connected neurons.
Moreover, the model is a generalization of the Boltzmann machine, extensively used in machine learning applications \cite{ackley_learning_1985}.
A popular application of the model involves inference of the model parameters to capture the properties of observed data. This inference process is referred to as the inverse Ising problem, where
kinetic Ising models \cite{witoelar_neural_2011} and their equilibrium counterparts \cite{schneidman_weak_2006} are used for modelling and analyzing biological systems. 

Unfortunately, exact solutions for describing network dynamics and inference often become computationally too expensive due to  combinatorial explosion of patterns in large systems, limiting applications using sampling methods to around a hundred of neurons \cite{tyrcha_effect_2013,tkacik_thermodynamics_2015}.
In consequence, analytical approximation methods are necessary for large networks. To this end, mean-field methods are powerful tools to track down otherwise intractable statistical quantities.

The standard mean-field approximations to study equilibrium Ising models are the classical naive mean-field (nMF) and  the more accurate Thouless-Anderson-Palmer (TAP) approximations \cite{thouless_solution_1977}.
In non-equilibrium networks, however, the system free energy is not directly defined, and it is not obvious how to apply mean-field methods. Alternatives involve the use of information geometric approaches \cite{kappen_mean_2000,aguilera2021unifying} or Gaussian approximations of the network effective fields \cite{mezard2011exact,mahmoudi_generalized_2014}. In this work, we will expand Gaussian mean-field approximations to explicitly address fluctuations for network simulation and inference. 

\section{Kinetic Ising model}

We model neural network activation using a kinetic Ising model, i.e. a  generalized linear model with binary states and pairwise couplings. The network consists of a system of $N$ interacting neurons $\*s_t$ (also called spins). The value of neuron $i$ at a time $t$ can take on two values $s_{i,t} \in \{+1,-1\}, i=1,2,\dots,N, t=0,1,\dots,T$ depending on the neuron being active or not.
At time $t$, the activation probability is defined by a nonlinear sigmoid function,
\begin{equation}
    P(s_{i,t} | \*s_{t-1}) = \frac{e^{s_{i,t}h_{i,t}}}{2 \, cosh \, h_{i,t}}.
    \label{eq:glauber-dynamics}
\end{equation}
Activation is driven by effective fields $\*h_{t}$, composed of a bias term $\*H = \{H_i\}$ and couplings to units at the previous time step $\*J= \{J_{ij}\}$,
\begin{equation}
\label{eq:h}
    h_{i,t} = H_i + \sum_j J_{ij}s_{j,t-1}.
\end{equation}
When the couplings are asymmetric (i.e, $J_{ij} \neq J_{ji}$), the system is away from equilibrium because the process is irreversible with respect to time.

In this article, we are interested in estimating first and second-order statistical moments of the system. That is, the components of the mean activation of a system and the fluctuations around this mean. Thus, we will calculate the activation rates $\*m_{t}$, correlations between pairs of units (covariance function) $\*C_{t}$, and delayed correlations $\*D_{t}$ defined as
\begin{align}
    m_{i,t} = &\sum_{\*{s}_{t}} s_{i,t}P(\*{s}_{t}),
    \label{eq:mean}
    \\C_{ik,t} = &\sum_{\*{s}_{t}} s_{i,t}s_{k,t}P(\*{s}_{t}) - m_{i,t} m_{k,t},
    \label{eq:correlation}
    \\D_{il,t} = &\sum_{\*{s}_{t},\*{s}_{t-1}} s_{i,t} s_{l, t-1} P(\*{s}_{t},\*{s}_{t-1}) - m_{i,t}m_{l, t-1}.
    \label{eq:D_correlation}
\end{align}

\section{Gaussian mean-field method}

In \cite{mezard2011exact}, the authors proposed that, in some cases, the second term of Eq.~\ref{eq:h} is a sum of a large number of weakly coupled components. Assuming weak and asymmetric couplings, given the Central Limit Theorem, they approximate this term by a Gaussian distribution $P(h_{i,t}) \approx \mathcal{N}(g_{i,t},\Delta_{i,t})$, with mean and variance:
\begin{align}
    g_{i,t} =& H_i + \sum_j J_{ij}m_{j,t-1},
    \\ \Delta_{i,t} =&  \sum_j J_{ij}^2 (1-m_{j,t-1}^2).
\end{align}
Yielding mean-field activation of neuron $s_i$ at time $t$ as:
\begin{equation}
   m_{i,t}  \approx \int D_z \tanh(g_{i,t} + z \Delta_{i,t} ),
\end{equation}
where  $D_z = \frac{dz}{\sqrt{2\pi}} exp(-\frac{1}{2}z^2)$ describes a Gaussian integral term with mean zero and unity variance. As well, the method provides a relation between $\*D_{t}$ and $\*C_{t-1}$. \cite{mezard2011exact}.

Alternatively, these equations can be derived by defining a mean-field problem using path integral methods  \cite{bachschmid-romano_variational_2016} or information geometry \cite{aguilera2021unifying}.
This approximation is exact in the thermodynamic limit for fully asymmetric methods \cite{mezard2011exact}.
However, in \cite{aguilera2021unifying} it was shown that this method (extended to add calculations of same-time correlations) fails to approximate the behaviour of fluctuations near a ferromagnetic phase transition for networks of hundreds of neurons.

\section{Conditional Gaussian mean-field method}

The motivation of this article is to explore extensions of the Gaussian mean-field method  in \cite{mezard2011exact} to accurately capture fluctuations in non-equilibrium systems, even in the proximity of critical dynamical regimes.

\subsection{Time-delayed correlations}
In order to better estimate correlations, instead of using a Gaussian approximation to compute $m_{i,t}$ (which results in a fully independent model), we propose the use of multiple conditional Gaussian distributions, aimed to capture conditional averages $m_{i,t}(s_{l,t-1})$ for a fixed neuron $l$ at the previous time-step $s_{l,t-1}$.
This conditional average can be approximated using a similar mean-field assumption:
\begin{align}
    m_{i,t}(s_{l,t-1}) =& \sum_{s_{t-1}} \tanh(h_i)P(s_{t-1}|s_{l,t-1}) \nonumber \\
    \approx &  \int D_z \tanh( g_{i,t}(s_{l,t-1}) +z \Delta_{i,t}(s_{l,t-1}) ),
    \label{eq:mf_cms}
\end{align}
where the statistical moments of the Gaussian distribution are computed as
\begin{align}
    g_{i,t}(s_{l,t-1}) =& H_i + \sum_j J_{ij}m_{j,t-1}(s_{l,t-1}), \\
    \Delta_{i,t}(s_{l,t-1}) =&  \sum_j J_{ij}^2 (1-m_{j,t-1}^2(s_{l,t-1})).
\end{align}
Here, $m_{j,t-1}(s_{l,t-1})$ are now conditional averages of two spins at time $t-1$. As a pairwise distribution $P(s_{j,t-1},s_{l,t-1})$ is completely determined by its moments $ m_{j,t-1},  m_{l,t-1}, C_{jl,t-1} $, we derive the equivalence
\begin{align}
    m_{j,t-1}(s_{l,t-1}) =& \sum_{s_{j,t-1}} s_{j,t-1} P(s_{j,t-1}|s_{l,t-1})
    \nonumber\\=& m_{j,t-1} + \frac{s_{l,t-1}-m_{l,t-1}}{1-m_{l,t-1}^2}C_{jl,t-1}.
\label{eq:m_j_s_l}
\end{align}

Once $m_{i,t}(s_{l,t-1})$ is known (Eq.~\ref{eq:mf_cms}), computing the marginal over $s_{l,t-1}\in \{1,-1\}$ we calculate $m_{i,t}$ as
\begin{align}
    m_{i,t} =& \sum_{s_{l,t-1}} m_{i,t}(s_{l,t-1})P(s_{l,t-1}) 
    = \sum_{s_{l,t-1}}m_{i,t}(s_{l,t-1} ) \frac{1 + s_{l,t-1}m_{l,t-1}}{2}.
    \label{eq:exact_m_i}
\end{align}

Finally, having the values of the conditional magnetizations we compute time-delayed correlations $D_{il,t}$ as
\begin{align}
    D_{il,t} =& \sum_{s_{i,t} s_{l, t-1}}  s_{i,t} s_{l,t-1} P(s_{i,t}, s_{l,t-1}) - m_{i,t} m_{l,t-1}
   \nonumber\\ =&  \sum_{s_{l,t-1}} m_{i,t}(s_{l,t-1})  \frac{s_{l,t-1} + m_{l,t-1}}{2} - m_{i,t} m_{l,t-1}.
   \label{eq:D_mi_sl2}
\end{align}

This sequence approximates the values of $\*m_{t}, \*D_{t}$ knowing the values of $\*C_{t-1}$. In order to recursively apply this method, we need to complement our equations with a method for computing $\*C_{t}$ from  $\*m_{t}, \*D_{t}$.

\subsection{Equal-time correlations}

We follow a similar procedure to approximate equal-time correlations. First, we  calculate the conditional average $m_{i,t}(s_{k,t})$, now conditioned on a neuron at the same time:
\begin{align}
    m_{i,t}(s_{k,t}) =& \sum_{\*s_{t-1}} \tanh(h_i)P(\*s_{t-1}|s_{k,t}) \\
    \nonumber \approx &  \int D_z \tanh( g_{i,t}(s_{k,t}) + z\Delta_{i,t}(s_{k,t})) ,
\end{align}
with moments
\begin{align}
    g_{i,t}(s_{k,t}) =& H_i + \sum_j J_{ij}m_{j,t-1}(s_{k,t}),
    \\ \Delta_{i,t}(s_{k,t}) =&  \sum_j J_{ij}^2 (1-m_{j,t-1}^2(s_{k,t})).
\end{align}
Here, we see that the Gaussian integral depends on averages $m_{j,t-1}(s_{k,t})$, conditioned on the next time step. We determine these quantities from the delayed correlations computed by Eq.~\ref{eq:D_mi_sl2} at the previous step
\begin{equation}
    m_{j,t-1}(s_{k,t}) = m_{j,t-1} + \frac{s_{k,t}-m_{k,t}}{1-m_{k,t}^2}D_{kj,t}.
\end{equation}

Once computed this conditional magnetization value, and having obtained the magnetizations from Eq.~\ref{eq:exact_m_i},  correlations are computed as:
\begin{align}
    C_{ik,t} =& 
    \sum_{ s_{k,t} } 
    m_{i,t}(s_{k,t})  \frac{s_{k,t} + m_{k,t}}{2} - m_{i,t} m_{k,t}.
\end{align}

\section{Results}

In this section, we compare the performance of our method with respect of two widely used methods: the TAP equations \cite{roudi_dynamical_2011} and the Gaussian mean-field method \cite{mezard2011exact} (implemented as in \cite{aguilera2021unifying} to account for same-time correlations). 
We test the methods 1) in an asymmetric version of the well-known Sherrington-Kirkpatrick (SK) model, and 2) in vitro recordings of neuron spiking activity from the mouse somatosensory cortex  \cite{ito2016spontaneous}. 

\subsection{Sherrington-Kirkpatrick model}

To test the methods, we use a dataset with simulations of an asymmetric kinetic version of the SK model with $N=512$ neurons \cite{aguilera_unifying_2020-dataset}. 
The asymmetrical SK model is known to have a phase transition between an ordered an disordered phases \cite{aguilera2021unifying} (although the spin glass phase is absent for fully asymmetric models). This critical point maximizes fluctuations of the system, thus being challenging for mean-field methods for finite sizes. 
Approximating network behaviour near criticality is highly relevant as  many biological systems, like neural networks, are believed to be poised near critical points \cite{tkacik_thermodynamics_2015}.

External fields $\*H_{i}$ are sampled from independent uniform distributions $\mathcal{U}(-\beta H_0, \beta H_0)$, $H_0=0.5$, whereas coupling terms $J_{ij}$ are sampled from independent Gaussian distributions $\mathcal{N}(\beta \frac{J_0}{N},\beta^2 \frac{J_\sigma^2}{N})$, $J_0=1, J_\sigma = 0.1$, where $\beta$ is a scaling parameter (i.e., an inverse temperature). The model displays a ferromagnetic phase transition, which takes place at $\beta_c \approx 1.1108$ \cite{aguilera2021unifying}.

\subsubsection{Network dynamics}
\label{sub:forward}

\begin{figure*}[t]
\centering{
\includegraphics[width=\linewidth]{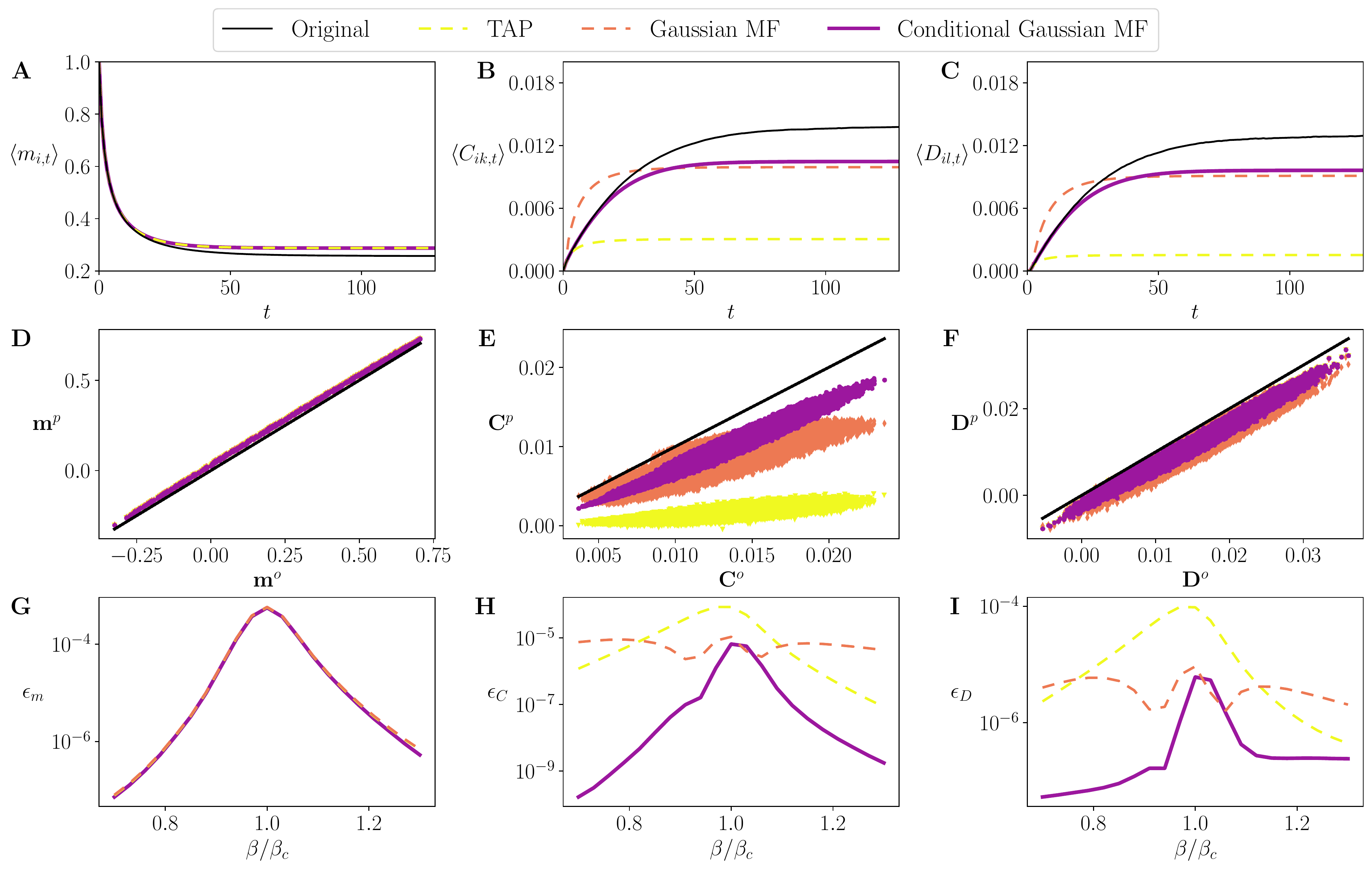}
}
\caption{\textbf{Approximation of neural dynamics in the SK model}. 
Top: Evolution of average magnetizations (A), equal-time correlations (B) and delayed correlations (C) found by different mean-field methods for $\beta=\beta_c$.
Middle: Comparison of magnetizations (D), equal-time correlations (E)  and delayed correlations (F) found by the different mean-field approximations (ordinate, $p$ superscript) with the original values (abscissa, $o$ superscript) for $\beta=\beta_c$ and $t=128$. Black lines represent the identity line.
Bottom: Mean Squared Error (MSE) of the magnetizations $\epsilon_{\*{m}}= \langle \langle (m^o_{i,t} - m^p_{i,t})^2 \rangle_{i} \rangle_t$ (G), equal-time correlations $\epsilon_{\*{C}}= \langle \langle (C^o_{ik,t} - C^p_{ik,t})^2 \rangle_{ik} \rangle_t $ (H), and delayed correlations $\epsilon_{\*{D}}= \langle \langle (D^o_{ik,t} - D^p_{ik,t})^2 \rangle_{il} \rangle_t$ (I) for 21 values of $\beta$ in the range $[0.7\beta_c, 1.3\beta_c]$.
} 
\label{fig:results-forward}
\end{figure*}

First, we examine the performance of the different methods at the time of computing the statistics of the model, i.e.,  $\*m$, $\*C$, and $\*D$. To this end, we start from a SK model with $\*s_0 = \*1$. We simulate the behaviour of the model  for $T=128$ steps comparing exact and mean-field behaviour for different values of the inverse temperature $\beta \in [0.7\beta_c, 1.3\beta_c]$.
In Fig.~\ref{fig:results-forward}A,B,C we observe respectively the average evolution of the magnetizations and equal-time and delayed correlations from time $t=0$ to time $t=128$. Fig.~\ref{fig:results-forward}D,E,F shows also a direct comparison between approximated and real values. We observe that our method makes the best approximation at the critical point. The approximations of both  $\*m$ and $\*D$ are very close to the identity line, however, small errors are accumulated resulting in a less accurate prediction of $\*C$. Besides, Fig.~\ref{fig:results-forward}G,H,I shows that our method performs better than the others at all inverse temperatures, including near the critical point.

\subsubsection{Inference}
\label{sec:inverse}

\begin{figure*}[htbp]
\centering{
 \includegraphics[width=0.7\linewidth]{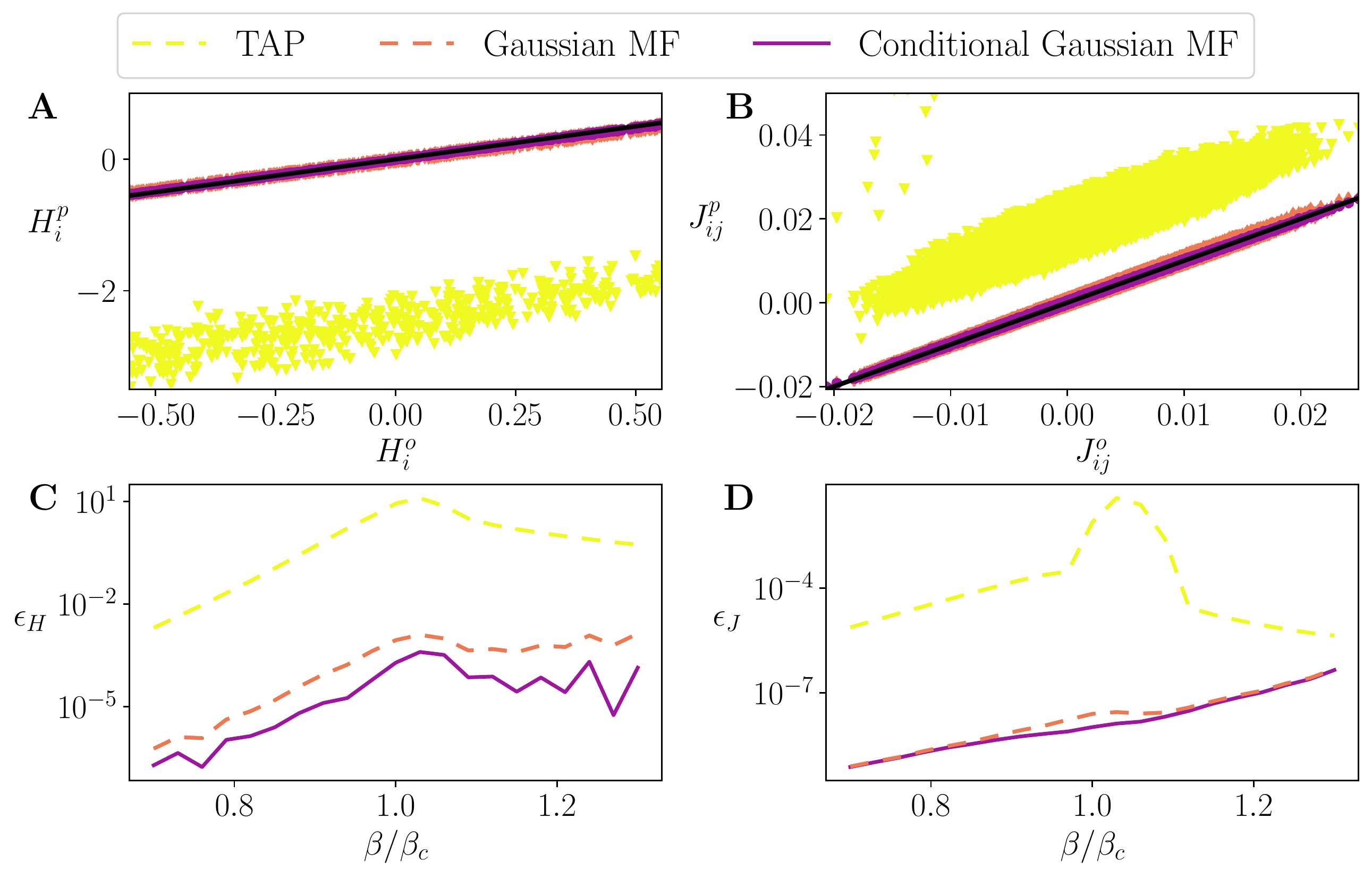}
}
\caption{\textbf{Network inference in the SK model}. Top: Inferred external fields (A)  and couplings (B) found by different mean-field models, plotted versus the real ones for $\beta=\beta_c$. Black lines represent the identity line. Bottom: Mean Squared Error of inferred external fields $\epsilon_{\*{H}}= \langle (H^o_{i} - H^p_{i})^2 \rangle_{i}$ (C) and couplings $\epsilon_{\*{J}}= \langle (J^o_{ij} - J^p_{ij})^2 \rangle_{ij}$ (D) for 21 values of $\beta$ in the range $[0.7\beta_c, 1.3\beta_c]$.
}
\label{fig:results-inverse}
\end{figure*}

Second, we compare the performance of the different methods in the inverse Ising problem, i.e., in inferring the model parameters from data. Starting from $\*H=0$ and  $\*J=0$, a gradient ascent on these parameters is performed using the maximum log-likelihood Boltzmann learning rule by means of approximating $\*m$ and $\*D$ using the mean-field equations (see \cite{aguilera2021unifying}). The Boltzmann learning algorithm is run a maximum of $R=10^6$ trials per step.
Fig.~\ref{fig:results-inverse}A,B displays the inferred local external fields ($\*H$) and couplings ($\*J$) of the model plotted against the real ones. We observe how the results for the TAP are displaced away from the identity line and how the results for the Gaussian mean-field method and ours are very similar. However, in Fig.\ref{fig:results-inverse}C,D we observe that our method obtains a lower Mean Squared Error ($\varepsilon$) for all inverse temperatures. We also observe that the error is lower at predicting couplings, which could be useful for studying the topology of biological neural circuits in setups more challenging to learn.

\subsubsection{Phase transition reconstruction}

Finally, we reconstruct a phase transition in the model by combining the inverse and forward Ising problem. As we know of the existence of a critical phase transition at $\beta_c$, we are interested in knowing how the different methods reconstruct the statistics around the phase transition point.
We use the $\*H$ and $\*J$ inferred in the inverse problem to calculate the systems' statistical moments and determine if the learned model is able to reproduce the original behaviour of the system in unobserved conditions. 
In order to reproduce the phase transition, the learned $\*H$ and $\*J$ are multiplied by a fictitious temperature in the range $[0.7\beta_c, 1.3\beta_c]$.

\begin{figure*}[h]
\centering{
\includegraphics[width=\linewidth]{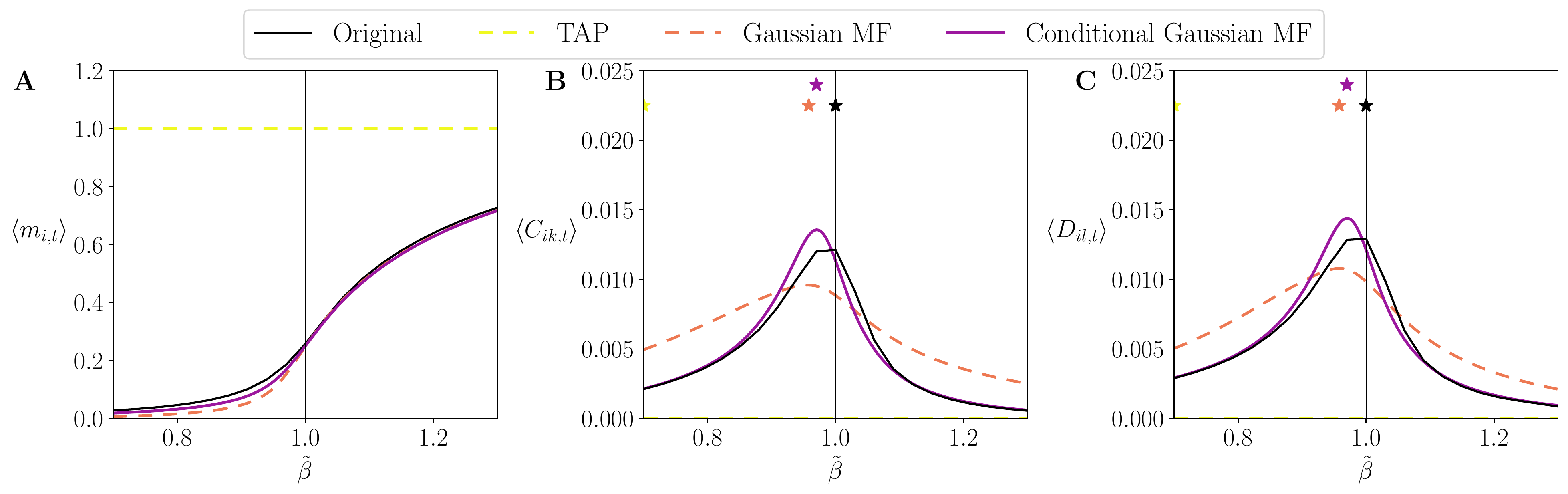}
}
\caption{\textbf{Phase transition reconstruction of the SK model}. 
Average of the Ising model's magnetizations (A), non-diagonal equal-time correlations (B), and  non-diagonal delayed correlations (C), at the last step $t=128$ of a simulation, found by different mean-field methods using the reconstructed network $\*H, \*J$ by solving the inverse Ising problem at $\beta=\beta_c$ and multiplying the estimated parameters by a fictitious inverse temperature $\tilde \beta$. The stars indicate the values of $\tilde \beta$ with maximum fluctuations.} 
\label{fig:results-reconstruction}
\end{figure*}

In Fig.~\ref{fig:results-reconstruction}A,B,C we display the averaged statistical moments after application of the inverse-forward pass. We observe how our method outperforms the others, achieving a better adjustment of the average of the statistical moments. While the Gaussian mean-field method from \cite{mezard2011exact} achieves a good approximation far from the critical point, our method achieves a close approximation at all inverse temperatures. 
Our method not only reduces the difference between the approximation and the expected values for the system's statistics, but it also preserves the shape around the critical point where other methods flatten out. This is of great interest because our method could help to better characterize non-equilibrium systems of high dimensionality poised near a phase transition.

\subsection{In vitro neuronal spike train data}

Finally,  we test the performance of our conditional Gaussian mean-field method on in vitro neural dynamics. To this end, we  selected a dataset containing neural spiking activity from mouse somatosensory cortex in organotypic slice cultures \cite{ito2016spontaneous}. Additional information about this dataset can be consulted in \cite{ito2014large-scale,alan2003what}. Specifically, we selected dataset 1, which contained 166 neurons. 


In order to adjust the dynamics of the dataset and our parallel update Ising model, we binned spike ocurrences in discrete time windows of length $\delta t$ and we extended the model to introduce asynchronous updates \cite{AsyncIsing} in which each neuron $s_{i,t}$ is updated with Eq.~\ref{eq:glauber-dynamics} with a probability $\gamma$, and updated to its previous value $s_{i,t-1}$ otherwise. From this data, we calculate the statistical moments $\*m,\*C,\*D$.

\subsubsection{Inference and network dynamics}

We infer the model parameters that best fit the data applying the Boltzmann learning algorithm starting from $\*H=0$ and $\*J=0$. After learning the model, the network is simulated for 128 time steps to reach a steady state. Different learning hyper-parameters were manually selected and tested, resulting in $\delta t = 70ms$ and $\gamma=0.77$. Early stopping was used, resulting in $1050$ iterations, taking the minimum MSE of one-step-estimation of $\*C$ as the stopping criterion.

\begin{figure*}[t]
\centering{
\includegraphics[width=\linewidth]{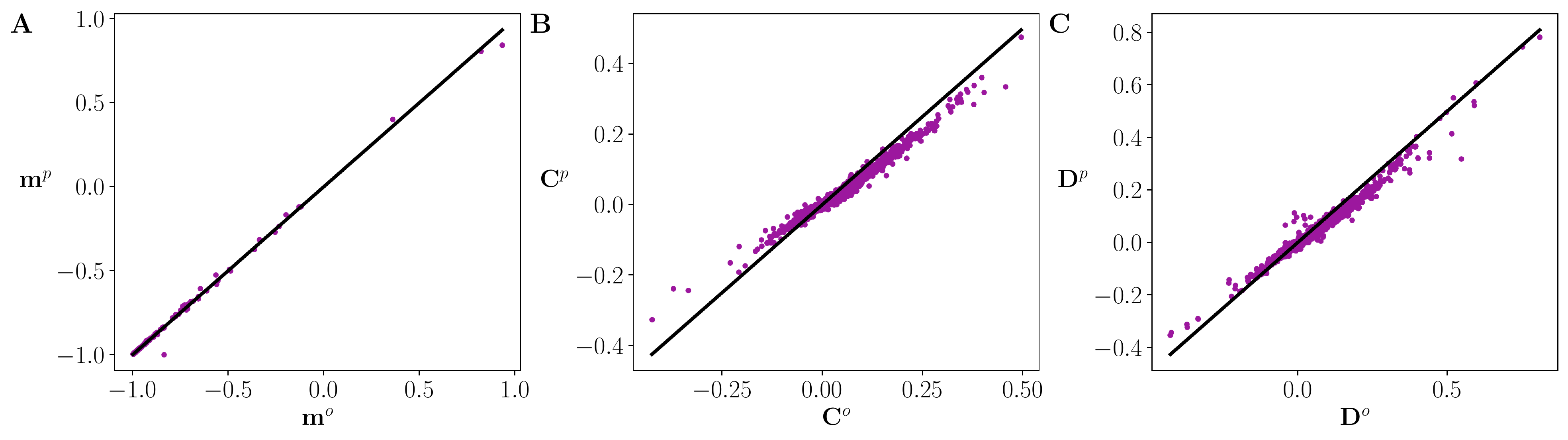}
}
\caption{\textbf{Inference and neural dynamics approximation of in vivo neural observations}. 
Comparison of magnetizations (A), equal-time correlations (B)  and delayed correlations (C) found after solving the inverse Ising problem (ordinate, $p$ superscript) with the original values (abscissa, $o$ superscript) for $\beta=1.0$ and $t=128$. Black lines represent the identity line.
} 
\label{fig:neurons-simulation}
\end{figure*}

After learning, we generated new data simulating the corresponding kinetic Ising model with the inferred parameters with asynchronous updates with probability $\gamma$.
In Fig.~\ref{fig:neurons-simulation} we compare the  statistics of the inferred model with respect to the original values. As we observe from the figure, almost all the statistics for each neuron lie near the identity line, leading to a MSE of $\epsilon_{\*{m}} =1.18\mathrm{e}{-06}$, $\epsilon_{\*{C}}=6.96\mathrm{e}{-06}$  and $\epsilon_{\*{D}}=6.26\mathrm{e}{-06}$ at the last step of the simulation. 

\subsubsection{Phase transition reconstruction}

\begin{figure*}[h]
\centering{
\includegraphics[width=\linewidth]{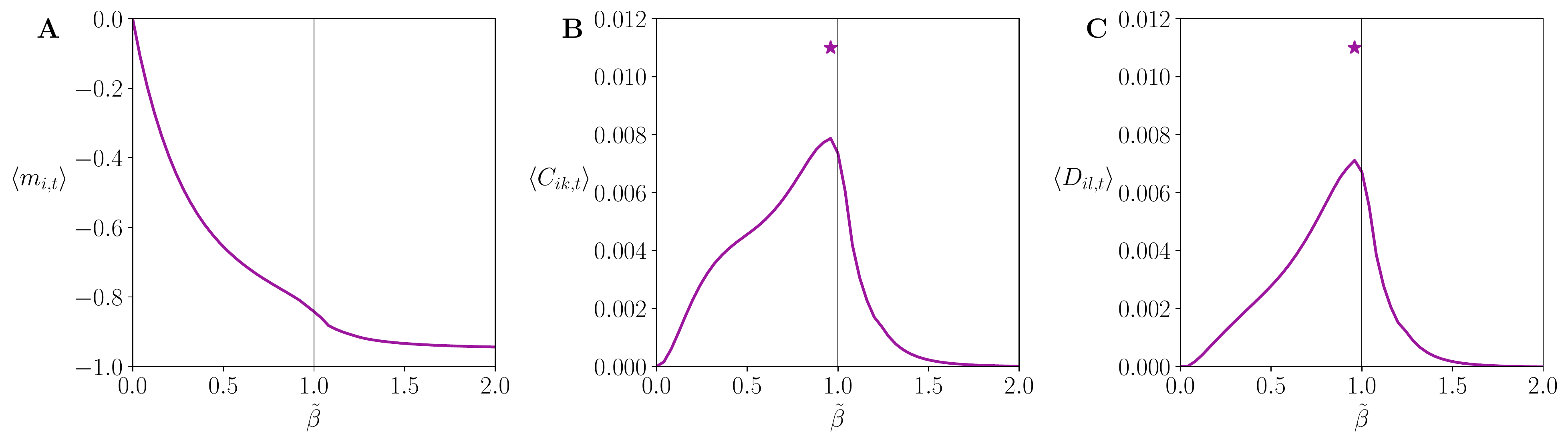}
}
\caption{\textbf{Phase transition of in vivo neural observations}. 
Average of the Ising model's magnetizations (A), non-diagonal equal-time correlations (B), and non-diagonal delayed correlations (C), by solving the inverse Ising problem at $\beta=1.0$ and multiplying a fictitious inverse temperature $\tilde \beta$ to the estimated parameters at  time $t=128$.
The stars are marked at the values of $\tilde \beta$ that yield maximum fluctuations.} 
\label{fig:neurons-reconstruction}
\end{figure*}

Finally, we explore if the model inferred from the data presents signatures of a similar phase transition that the asymmetric SK model. We multiply  $\*H$ and $\*J$ by a fictive inverse temperature $\tilde \beta$ in the range $[0, 2]$. After this, we simulate every neural system for $T=128$ steps. Fig.~\ref{fig:neurons-reconstruction} displays the average statistics at the different  temperatures. Again,  as in the SK model (Fig.~\ref{fig:results-reconstruction}) we observe a peak in correlations around the operating temperature (i.e. $\tilde \beta =1$), suggesting the presence of a continuous phase transition. Further analysis (e.g. testing of different network sizes, entropy estimations \cite{tkacik_searching_2014}) should be performed to confirm this result.

\section{Discussion}

Many biological networks are found to self-organize at points of their parameter space, maximizing fluctuations in the system \cite{mora_dynamical_2015} and showing non-equilibrium dynamics. Although mean-field methods have been successfully proposed as a tool to approximate complex network phenomena and transitions, successfully capturing fluctuations in non-equilibrium conditions is a challenging open problem.
Here, we extend a previous method in the literature describing a Gaussian mean-field estimation of the average activation of a system \cite{mezard2011exact}. This method is known to accurately capture average activations in fully asymmetric networks, but capturing fluctuations or transitions in networks presenting different degrees of symmetry is still challenging.
We have shown how an extension based in computing  Gaussian estimations of the conditional input field offers a good approximation of pairwise correlations even in the proximity of a ferromagnetic phase transition.
This is specially important as it allows not only to simulate network dynamics taking into account fluctuations, but also to offer fast methods to do inference in neural networks (i.e. solving the inverse Ising problem or the equivalent Boltzmann learning problem).

Our results show how these methods present a good performance in well-known neural network theoretical models. As well, we show in a preliminary test how the method is able to successfully infer a model reproducing the statistics of neural spike trains recorded form a sensorimotor cortex culture, suggesting that it operates near a critical phase transition.  This is expected to foster useful tools to efficiently analyse large-scale properties of neural network dynamics.

\section*{Acknowledgements}
M.A. was funded by the European Union's Horizon 2020 research and innovation programme under the Marie Skłodowska-Curie grant agreement No \mbox{892715}.
%
%
%
\bibliographystyle{splncs04}
\bibliography{main}

\end{document}